# Rethinking Interphase Representations for Modeling Viscoelastic Properties for Polymer Nanocomposites

Xiaolin Li[1], Min Zhang[2], Yixing Wang[3], Min Zhang[4], Aditya Prasad[5], Wei Chen[3], Linda Schadler[6], L. Catherine Brinson[2,*]

[1]Theoretical and Applied Mechanics, Northwestern University, Evanston, IL, 60208

[2]Mechanical Engineering and Materials Science, Duke University, Durham, NC, 27708

[3]Mechanical Engineering, Northwestern University, Evanston, IL, 60208

[4]Materials Science and Engineering, Northwestern University, Evanston, IL, 60208

[5]Materials Science and Engineering, Rensselaer Polytechnic Institute, Troy, NY, 12180

[6]College of Engineering and Mathematical Sciences, the University of Vermont, Burlington, VT 05405

[*]Corresponding author, contact: cate.brinson@duke.edu

**Abstract**

Numerical modeling of viscoelastic properties is critical to developing the structure-property relationship of polymer nanocomposites. While it is recognized that the altered polymer region near filler particles, the interphase, significantly contributes to enhancements of composite properties, the spatial distribution of interphase properties is rarely considered due to lack of local property measurements. In recent years, the Atomic Force Microscopy (AFM) technique has begun to make local property measurements of the interphase available. In the light of the increasing availability of AFM data, in this work a new interphase representation for modeling the viscoelastic properties of polymer nanocomposites is proposed. The proposed interphase representation disentangles the interphase behavior by two separate components – *single-body interphase gradient* and *multi-body compound effect*, whose functional forms are learned via data mining. The proposed interphase representation is integrated into a Finite Element model, and the effects of each component of the interphase representations are numerically studied. In addition, the advantages of the proposed interphase representation are demonstrated by comparison to a prior simulation work in which only a uniform effective property of the interphase is considered. Moreover, the proposed interphase representation is utilized to solve the inverse problem of inferring spatial distribution of local properties using Bayesian Inference on experimental data.

## 1. Introduction

Polymer nanocomposites have attracted much attention in materials science due to their superior properties and great potential in engineering applications. Prior studies [1-8] have shown that in polymer nanocomposites, the incorporation of a small amount of nanoscale inclusions renders significant enhancements of their mechanical, dielectric and thermal properties, while retaining low density and ease of processing. This combination of multi-property enhancement and facile manufacture make polymer nancomposites a promising multi-functional material in industry. A variety of constituents (e.g. silica[9], clay[10] and graphite[11]) and morphologies (e.g. spherical particle[12] or nanotube[13]) of nano-inclusions have been investigated to meet different engineering demands in prior works.

In addition to attributes offered to the composite by the nano-inclusions themselves, the region of polymer near the particles, termed the interphase, contributes dramatically to the overall material response. Both geometric and chemical constraints at the nanofiller-polymer interface alter the mobility of the local and interconnected polymer chains, resulting in significant changes to the physical properties of this interphase domain[14-17]. Most notably, even in nanocomposites with very low loadings (e.g. 1 wt % filler), the percolating nature of polymer chain response can lead to a substantial interphase volume fraction, and thereby contribute to the bulk composite properties. For example, it is demonstrated in [12] that the existence of ~2 vol% well dispersed nanotubes will produce 81.4 vol% interphase areas assuming an interphase thickness 3 times the diameter of nanotube.

In recent years, experimental efforts have been devoted to studying effects of interphases on the bulk properties of polymer nanocomposites. For instance, Seiler et al. [14] applied Electric Force Microscopy (EFM) to verify the existence and to estimate the thickness of interphase in silicon nanocomposites. Ciprari et al. [18] utilized data from thermal gravimetric analysis (TGA), transmission electron microscopy (SEM) and Fourier transform infrared spectroscopy (FTIR) to study the structure and density of interphase for six nanocomposite systems. In addition to such experimental investigations, theoretical models have also been proposed to analyze the impacts of interphase for composites. Review articles by Hashin[19] and Christensen[20] cover fundamental mathematical models developed in early years for composite material behaviors without considering interphase, including the popular Mori-Tanaka method[21]. Fisher and Brinson[22] further improve the combination of the Mori-Tanaka method and Benveniste's method[23] by assuming an interphase region between filler aggregates and polymer to investigate the mechanical property of viscoelastic composite. Other analytical approaches for analytically modeling the interphase includes Deng et al. [24], in which the particle-interphase regions are mechanically equated to effective particles by applying a volume fraction weighted super-position of particles and matrix, and Ji et al. [25] where a simple linear gradient change of elastic moduli in the interphase was assumed. While having demonstrated the success in evaluating the structure-property relationship of polymer nanocomposites, the existing models are limited by only considering the microstructure in a coarse or average sense.

Promoted by the developments of modern computational methods, Finite Element (FE) modeling provides an inexpensive alternative to the traditional experimental approaches and micromechanics models. A number of works have utilized FE methods to study the effects of interphase in composites. For instance, Zhu et al. [27] developed a FE model to predict the elastic properties of polymer nanocomposites and showed good agreement with experimental measurements. In Boutaleb et al. [26], a micromechanical analytical interphase model, in which interphase regions are also assumed, is proposed for studying and modeling the stiffness of polymer composites via Finite Element Analysis. Another promising example is Read et al. [28], which presents a FE model for predicting the viscoelastic property of polymer blends whose polymer constituents are immiscible and the property distribution is discrete. Our earlier works [12, 29-32] also illustrated FE models to study effects of interphase on viscoelastic property of polymer nanocomposites. In one pair of papers, [12, 29] we presented a FE model in which uniform interphase properties are assumed, and the effects of interphase volume fraction and particle agglomeration, respectively, are explored. Built upon this uniform interphase FE model, studies regarding choice of Representative Volume Element (RVE) [30], statistical assembly of Statistical Volume Elements [31] and inference of appropriate interphase parameters from experiments [32] have been pursued. Despite these successes, it is noteworthy that in these works [12, 29-32], the effective interphase property has been modeled by one or two uniformly distributed interphase layers surrounding the filler aggregates. This simplified assumption of the interphase domain is made due to lack of local measurements of interphase properties.

In recent years, the development of Atomic Force Microscopy (AFM) nano-indentation instrumentation has provided a feasible solution to measure the mechanical property of interphase domains in the nanoscale. In Downing et al.[33], it is illustrated that the size of the interphase and its stiffness could be measured using phase imaging AFM. In addition, Cheng et al.[34] demonstrated the feasibility of using AFM nano-indentation to characterize local elastic modulus on a film-substrate model composite. Zhang et al. [35] created a set of substrate-polymer-substrate samples, varying the distances between substrates, and discovered both the interphase gradient decay and interactions between interphases produced by different substrates. The availability of AFM data, together with the recent development of statistical tools for analyzing and quantifying microstructures [36], make it feasible to describe interphase behaviors continuously (in contrast to the discretized/uniform representations as used in prior works) and take the spatial effects of interphase and microstructural dispersions into account. In this regard, this work presents a new descriptive interphase representation for modeling the properties of polymer nanocomposites, with an explicit implementation for viscoelastic material response. The approach, however, is general and applicable to other physical properties. The proposed interphase representation disentangles the complex interphase behavior into two continuous interphase functions, namely the *single-body interphase gradient* and the *multi-body interphase compound effect*. The single-body interphase gradient describes the spatial distribution of the material property within the interphase created by one single filler aggregate, while the multi-body interphase compound effect quantifies the interacting behavior of interphases created by different aggregates. The functional formulations of these interphase functions are learned via mining a set of AFM data in [35], and they are implemented into a pixelated plain strain FE model. Numerical studies are conducted to investigate the effects of each interphase function on the bulk viscoelastic property of the polymer composites.

By a comparative study with prior simulation work [12, 29, 30], the advantage of the proposed interphase representation and the corresponding FE model is demonstrated.

The remainder of the paper is organized as follows: in Section 2, the proposed interphase representations are first defined. We then illustrate how the proposed interphase representation can be implemented in FE modeling of the viscoelastic properties of polymer nanocomposites. In addition, a Bayesian Inference (BI) approach, which is an essential tool for inferring the interphase representation mathematically, is introduced. In Section 3, the explicit functional forms of the interphase representation are identified via data mining, followed by a series of numerical studies to demonstrate the effects of the interphase representations. The advantages of the proposed interphase representation are then studied via a comparative study using BI with the prior interphase modeling work [12]. Moreover, by utilizing a set of experimental data, the spatial distribution of local viscoelastic property is investigated via solving the inverse problem by BI. In Section 4, we draw conclusions and discuss future work.

## 2. Method

### 2.1 Interphase Representation

While it is well recognized that the existence of interphase significantly contributes to the bulk viscoelastic properties of polymer nanocomposites, there is not a descriptive interphase representation that takes the spatial distribution of interphase properties into account. This deficiency is primarily because: 1). In early years, local measurements of the polymer properties at 10's of nanometer resolution were not available and thus there was insufficient experimental evidence to support a detailed interphase representation, and 2). The interphase effects on the polymer nanocomposites are a coupled interaction of both spatial property distribution and microstructural dispersions. Therefore, it is difficult to exclude the influence of complex microstructural dispersions and isolate the interphase spatial distribution alone. To address this entanglement, Cheng et al. [34] manufactured a special film-substrate structured "model" composite sample and utilized AFM to study the local property of interphase created by single filler aggregates. Zhang et al. [35] extended this work by producing a series of substrate-film-substrate model composite samples with different film thicknesses. These works eliminated the effects of microstructure dispersion and demonstrated that 1) the stiffness of the interphase decays as the distance from the filler interface increases. 2) closer pairs of substrates result in an increased stiffness in the interphase region, beyond simple superposition of two single-surface effects. In the light of these experimental observations, in this work, we propose a new gradient interphase representation for modeling the viscoelastic property of polymer nanocomposites. Specifically, two interphase functions, namely *single-body interphase gradient* and *multi-body interphase compound effect*, are proposed to reflect and quantify the experimental findings.

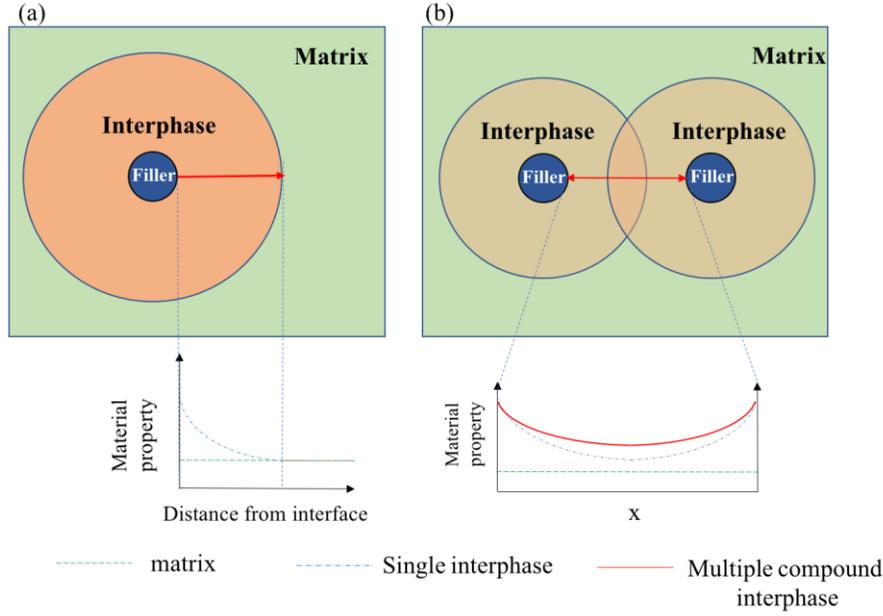

**Figure 1. The illustration of interphase functions. (a) Single-body interphase gradient. (b) Multi-body compound effect. The blue dashed line represents the single-body interphase gradient created by one single aggregate and the red curve shows how the compound effect makes an impact on the resultant interphase property.**

### 2.1.1 Single-body Interphase Gradient

It is found in Cheng et al.[34] and Zhang et al. [35] that the local elastic property of interphase decays as the distance from filler surface increases. Based on this observation, we propose the first component of interphase representation, namely *single-body interphase gradient*, to describe the dependence of interphase property on the distance from a single filler aggregate. As shown in **Figure 1**(a), the single-body interphase gradient describes the gradually decaying effects in the interphase created by one single filler aggregate. No interaction between interphases created by different aggregates are considered in the single-body interphase assumption. For the elastic property, the single-body interphase gradient is expressed as,

$$E_d = F(d; \Phi), \quad d \leq L_0 \tag{1}$$

where $E_d$ is the elastic modulus at a distance d from filler surface, $F(\cdot)$ is the functional form of the interphase gradient defined by a set of hyper-parameters $\mathbf{\Phi}$, and $L_0$ is the interphase thickness which is usually determined by some threshold value on the mechanical property enhancement. To identify the functional form of the single-body interphase gradient function for viscoelastic properties, while recent works [37, 38] have shown the feasibility of directly measuring the local viscoelasticity via AFM, the methodologies to capture quantitative viscoelasticity from AFM instrumentation is still under development. Additionally, in [39] the microstructural geometries used are too complex to disentangle the coupling effects of microstructure and interphase gradient. Therefore, the analogy between the elastic modulus and the magnitude of the complex modulus is utilized in this work to extend the interphase gradient from elasticity as in [35] to viscoelasticity. The functional form of the single-body interphase gradient function $F(\cdot)$ can be expressed as,

$$\overline{|E_{d,f}^*|} = \frac{|E_{d,f}^*|}{|E_{+\infty,f}^*|} = F(d, f; \mathbf{\Phi}), \quad d \leq L_0 \tag{1}$$

where $E_{d,f}^*$ is the complex modulus of the interphase at a distance $d$ from the filler aggregate, $|E_{d,f}^*|$ represents the magnitude of the complex modulus, and $\overline{|E_{d,f}^*|}$ indicates that modulus normalized by the magnitude of the complex modulus of the matrix $|E_{+\infty,f}^*|$. In addition, $f$ is the frequency at which the viscoelasticity is measured, and $L_0$ is the interphase thickness and $\mathbf{\Phi}$ is a set of hyper-parameters in the functional form of the interphase decay. For the examples in this work, elastic AFM data is used, such that the F determined is independent of frequency and the frequency dependence for $\overline{|E_{d,f}^*|}$ arises solely due to the matrix viscoelastic modulus. However, as reliable viscoelastic AFM data becomes available, the interphase gradient function F can be revised to contain frequency dependencies explicitly as indicated in Eqn. (2).

### 2.1.2 Multi-body Compound Effect

In defining the single-body interphase gradient, it is prescribed that only the interphase created by one single aggregate is considered. While it seems that this definition could quantify the experimental findings in [34], it would fail to describe the interacting effects of interphases observed by Zhang et al. [35]. The single body gradient would also fail to capture interaction effects in complex microstructure dispersions, where every particle has many close neighbors and thus many potential interaction effects. Therefore, we propose the second component of interphase representation, namely *multi-body (interphase) compound effect,* to describe the interaction phenomenon in the interphase areas that are affected by multiple filler aggregates. The multi-body interphase compound effect decays to the single body effect if the filler aggregates are sufficiently far from each other, and it occurs when the interphases created by different aggregates interact as depicted in **Figure 1**(b). The multi-body interphase compound effect is not limited to simply additive response and an extended interphase region for interacting particles can be larger than that for isolated particles, as observed in [35]. To model the viscoelasticity of the interphase for multi-body effects, the analogy between the elastic modulus as in [35] and the magnitude of the complex modulus introduced in Section 2.1.1 is utilized again and the form of the multi-body compound effect $G(\cdot)$ is expressed as,

$$|\overline{E^*_{x,f}}| = G(|\overline{E^*_{d_1,f}}|, |\overline{E^*_{d_2,f}}|, \dots, |\overline{E^*_{d_n,f}}|; \boldsymbol{\Omega}) \tag{2}$$

where $|\overline{E^*_{x,f}}|$ is the normalized magnitude of the complex modulus at location **x**, $f$ is the frequency at which the viscoelasticity is measured, $d_i (i = 1, 2, \dots, n)$ represents the nearest distances from location **x** to the nearby filler aggregate #i, $n$ is determined by a user-specified, sufficiently large cut-off distance, and $\boldsymbol{\Omega}$ is a set of hyper-parameters in the functional form of the compound effect.

## 2.2 Implementing the Proposed Interphase Representation in Finite Element Modeling

### 2.2.1 Finite Element Model

To examine the influences of the proposed interphase representation functions (i.e. single-body interphase gradient and multi-body compound effect), a pixelated viscoelastic 2D plain strain model is developed. Different from our prior work [12] in which a conforming mesh is utilized, the pixelated model in this work discretizes the microstructure with elements of rectangular shapes, which as a consequence avoids potential meshing errors and provides higher flexibility in assigning complex interphase property distributions. In addition, periodic boundary conditions are applied in assigning the interphases (shown in **Figure 2**) and constraining the boundary displacements as,

$$\begin{aligned}\boldsymbol{u}(X_1, 0) + \boldsymbol{U_2} &= \boldsymbol{u}(X_1, L) \\ \boldsymbol{u}(0, X_2) + \boldsymbol{U_1} &= \boldsymbol{u}(L, X_2)\end{aligned} \quad (3)$$

where L is the length of the square edge, $\boldsymbol{u}(i,j)$ is the displacement at location $(i,j)$, and $\boldsymbol{U_1}$ and $\boldsymbol{U_2}$ depend on the particular loading applied on the cell.

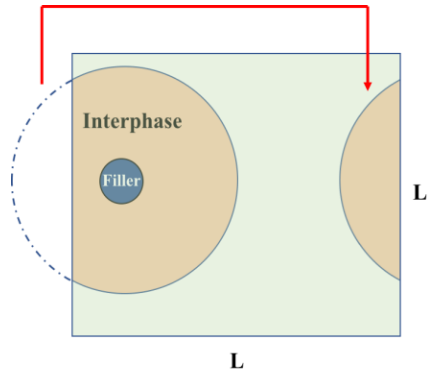

**Figure 2. The illustration of the periodic boundary assignments of interphase**

### 2.2.2 Interphase property

In prior numerical simulations for polymer nanocomposites [29, 30, 32, 40, 41], the effective property of the interphase is often assumed to be directly related to the property of the polymer matrix. Therefore, the interphase property is modeled by applying a transformation on the master curve of matrix. For instance, to simulate the dielectric property of polymer composites [41], the interphase property has been obtained through a transformation process controlled by five parameters, while to model the viscoelastic property[12, 30], a shifting and/or broadening conversion of the matrix property has been assumed. In this work, we assume that in the interphase created by a single filler aggregate, the local viscoelastic property of an infinitesimal area in the interphase is related to the bulk property by a shift of $S$ decades in relaxation time, and the shifting factor S is dependent on the distance away from the filler surface. For the interphase locations affected by multiple filler aggregates, the interphase property is determined using the multi-body compound effect to combine the theoretical single-body interphase properties. **Figure 3** illustrates these assumptions of interphase viscoelastic property conceptually.

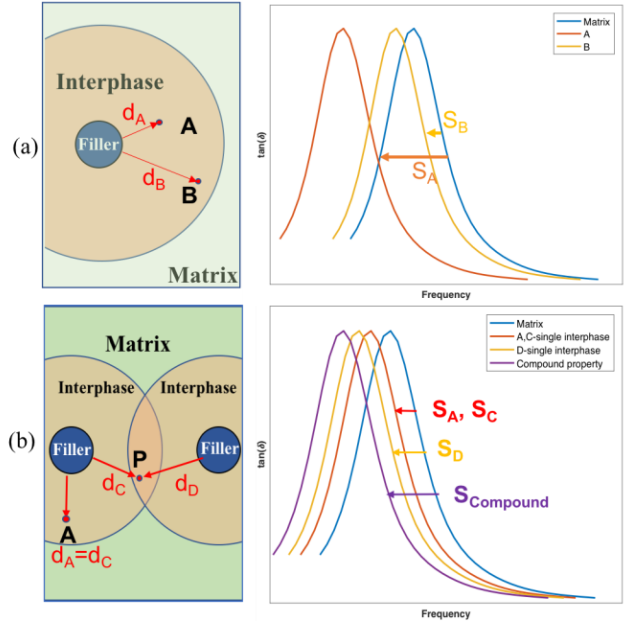

**Figure 3. The illustration of the assumptions of the interphase viscoelastic properties (a) left – the two locations A and B in the interphase are identified; right – the corresponding viscoelastic properties (tan$\delta$ peak) for A and B. Since A is closer to the filler than B, the shifting factor $S_A$ is greater than $S_B$. (b) left – location A again experience a single-body effect from the left filler, but point P, at the same distance from the left filler as point A, is located also close to the right filler in the interacting interphase region. For point P, the distances from two filler aggregates $d_C$ and $d_D$ are identified and used to determine the resultant property using single-body and multi-body effects; right – the "virtual" property for point P is found by first inferring single-body interphase gradient by $d_C$ ($S_C$) and $d_D$ ($S_D$). $S_{compound}$ for point P is then determined using the multi-body compound effect function $G(\cdot)$ in Eqn. 3 (detailed inference is illustrated in Section 3.2).**

It should be also noted that according to the guideline suggested by Cheng et al. [34], the interphase thickness, $L_0$, is defined as the distance from substrate surface to the point where the interphase property drops to 105% of the matrix property. This definition of interphase thickness $L_0$ is also utilized in this work when the interphase representation is implemented in FE model. In other words, a particular location is considered as interphase if its magnitude of the complex modulus is greater than 105% of that of the matrix. Using this criteria, locations which are outside the range of single-body interphase thickness from the aggregates can still be inside the range for the multi-body effect.

**2.3 Bayesian Inference (BI) for identifying the hyper-parameters in the interphase representation**

For polymer nanocomposites, techniques such as DMA for acquiring the bulk viscoelastic properties are comparatively mature and inexpensive while probing the local viscoelastic properties within the interphase using AFM is still under development, time consuming and relatively rare. Therefore, in this work, we apply a Bayesian Inference based approach [32] to inversely infer a reasonable distribution of properties within the interphase region. In this approach, AFM data on "model" composites is firstly utilized to learn the functional forms of the two interphase functions, in which the values of hyper-parameters $(\boldsymbol{\Phi}, \boldsymbol{\Omega})$ are unknown. The functional forms of the interphase representation are then implemented into the FE model and Bayesian Inference is conducted to identify the values of $(\boldsymbol{\Phi}, \boldsymbol{\Omega})$ that best match the simulated bulk composite property with the objective property. In a first demonstration, a virtual target property is computed using the prior uniform interphase model and is used as the objective to identify the equivalent gradient interphase representation. In the second demonstration, the gradient interphase modeling approach is applied to experimental data to investigate the spatial distribution of interphase properties. The first demonstration serves as a validation of the algorithms developed while the second demonstration shows the ability of the approach to describe experimental composite data with continuous interphase functions as developed in this paper.

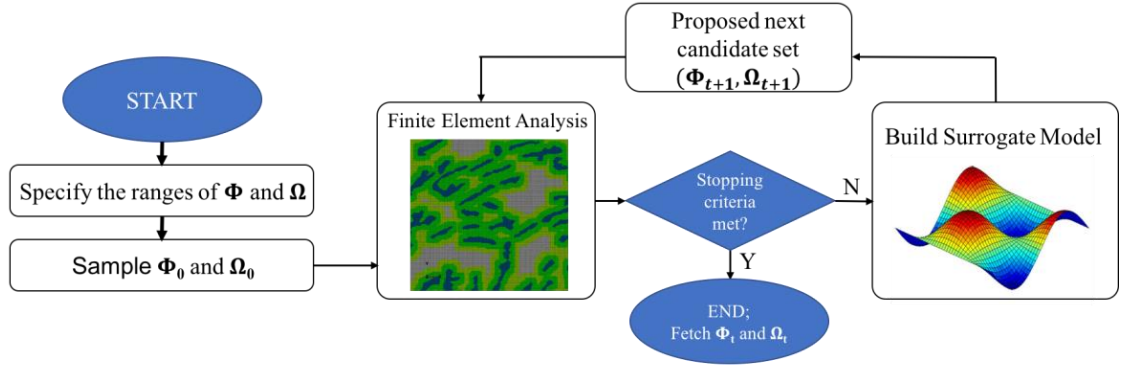

**Figure 4 The workflow of the Bayesian Inference approach for identifying the values of the hyper-parameters Φ and Ω.**

**Figure 4** illustrates the workflow of the Bayesian Inference approach to identify the hyper-parameters' values in the gradient interphase representation. Before the inference process, the properties of individual constituents (particle aggregates and polymer matrix), the bulk property of the composite and its microstructural dispersion is known, and the functional forms with unknown parameters $(\mathbf{\Phi}, \mathbf{\Omega})$ of the single-interphase gradient and multi-body interphase compound effect are identified via data mining on the AFM data. The objective of the Bayesian inference is to identify the appropriate values of $(\mathbf{\Phi}, \mathbf{\Omega})$ that match the simulated bulk composite property with experimental data. The Bayesian Inference starts with the specification of the ranges of the hyper-parameters $\mathbf{\Phi}$ and $\mathbf{\Omega}$, which in the context of interphase modeling confines the strength of single-interphase gradient property and interphase interactions. After that, the initial values $\mathbf{\Phi_0}$ and $\mathbf{\Omega_0}$, which determines the corresponding single-body interphase gradient and multi-body compound effect, are randomly generated from the prescribed ranges. The interphase representation is then fed into FE analysis to simulate the viscoelastic property. While it is theoretically possible to conduct a full grid search on all possible solutions $X = (\mathbf{\Phi}, \mathbf{\Omega})$ and evaluate the difference between the simulation property and the objective property (we denote the difference as $y$), it is computationally cheaper to infer the relationship of $X$ and $y$ via *Gaussian Process* metamodeling.

Gaussian Process model, also known as Kriging model, is a statistical model that interpolates the observations and supplies quantification of uncertainty for the metamodel prediction at each estimation point. Essentially, Gaussian Process models the data points $\{X, y\}$ and the estimations $\{X', y'\}$ using

$$\begin{bmatrix} y \\ y' \end{bmatrix} \sim \mathcal{N}\left(0, \begin{bmatrix} \text{Cov}(X, X) & \text{Cov}(X, X') \\ \text{Cov}(X', X) & \text{Cov}(X', X') \end{bmatrix}\right) \qquad (4)$$

in which $\text{Cov}(A, B)$ represents the covariance matrix between $A$ and $B$, defined by $\text{Cov}(A, B) = \mathbb{E}(AB^T) - \mathbb{E}(A)\mathbb{E}(B^T)$. Conditioning on the data $D = \{X, y\}$, the posterior $P(y'|X, X', y)$ yields a Gaussian distribution where,

$$\begin{aligned} \mu &= \text{Cov}(X, X')\text{Cov}(X, X')^{-1} y \\ \Sigma &= \text{Cov}(X', X') - \text{Cov}(X, X')\text{Cov}(X, X)^{-1}\text{Cov}(X', X) \end{aligned} \qquad (5)$$

Gaussian Process metamodeling establishes a surrogate model that quantifies the statistical mean and uncertainties in the unexplored region. By considering the mean estimation and the uncertainties, the next candidate point $X_{t+1}$ that could potentially improve the performance would be identified based on the current dataset $(X_{0:t}, y_{0:t})$. In proposing the next candidate point, several criteria have been used. For instance, [32] utilizes Expected Improvement (EI) while Li and Yang et al. [42, 43] applies the GP-Hedge criteria which combines three scores -- EI, lower confidence bound (LCB) and probability of improvement (PI). In this work, EI is utilized to propose the next candidate point $X_{t+1} = (\Phi_{t+1}, \Omega_{t+1})$ to explore.

## 3  Results

Utilizing the ideas and the related techniques presented in the prior section, the detailed analysis for identifying the functional forms of interphase gradients is demonstrated in this section. Specifically, the functional forms with unknown parameter set $(\boldsymbol{\Phi}, \boldsymbol{\Omega})$ of the interphase gradients are firstly explored via data mining on AFM data (Section 3.1). Then the correspondence between the interphase gradient and shifting factors in the frequency domain for viscoelasticity is illustrated in Section 3.2 to demonstrate how the proposed interphase gradient is implemented into Finite Element modeling. Lastly, Section 3.3 presents the numerical analysis of the impacts of $(\boldsymbol{\Phi}, \boldsymbol{\Omega})$ parameters (Section 3.3.1 and Section 3.3.2), as well as the numerical validations using Bayesian Inference by taking the prior uniform interphase model (Section 3.3.3) and an experimental data (Section 3.3.4) as optimization objectives.

### 3.1 Data mining to identify the functional forms of interphase representation

Experimental data on model samples provides clear, quantitative data on interphase gradients near surfaces [34, 35], and even illustrates a compound interaction effect. Herein we utilize the AFM data in [35] to identify appropriate functional forms of the interphase representation.

### 3.1.2 Single-body interphase gradient

Zhang et al.[35] present AFM data on modulus gradients on carefully designed model composites with variable spacing between substrates. Four PMMA samples with distances of 520nm, 256nm, 156nm and 60nm between the silica substrates were used. It is found that 520nm is sufficiently large to ensure that interphases created by the two substrates do not interact with each other. Therefore, the data of the 520nm sample is utilized to probe the functional form of single-body interphase gradient in this work. Data analysis reveals that the single-body interphase gradient follows an exponential decay and its functional forms could be expressed as:

$$\overline{E_d} = \alpha' e^{-\beta' d} + 1 \qquad (6)$$

where $\overline{E_d}$ is the normalized modulus, $\alpha'$ and $\beta'$ are hyper-parameters and $d$ is the distance from filler surface. The fits of the data are illustrated in **Fig. 5.**

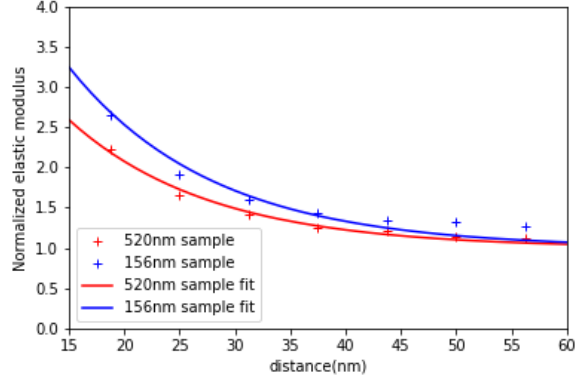

**Figure 5 Regressions of the AFM experimental data. The regression of 520 nm sample is utilized for investigating the functional form of the single-body interphase gradient** ($\alpha' = 5.14, \beta' = 0.079$)**, while the 156nm sample** ($\alpha' = 7.07, \beta' = 0.077$) **will be utilized later in the study of multi-body compound effect in Section 3.1.3.**

In order to extend these data on elastic interphase values into modeling the viscoelastic properties, we utilize the analogy between elastic modulus and the magnitude of complex modulus. The normalized magnitude of the complex modulus of the interphase in polymer nanocomposite is thus expressed as,

$$\overline{|E^*_{d,f}|} = \frac{|E^*_{d,f}|}{|E^*_{+\infty,f}|} = \alpha e^{-\beta d} + 1 \tag{7}$$

where $\overline{|E^*_{d,f}|}$ is the normalized magnitude of the complex modulus at distance $d$ from the filler under frequency $f$, and $|E^*_{+\infty,f}|$ is the magnitude of complex modulus of polymer matrix. It is also noted that in this extension from elasticity to viscoelasticity, it is assumed that the frequency dependence of the interphase property $E^*_{d,f}$ is identical to that of the matrix property $E^*_{+\infty,f}$. This functional form explicitly specifies the hyper-parameters defined in Eqn. 2 by $\mathbf{\Phi} = (\alpha, \beta)$. It should be noted that, while both Eqns. 7 and 8 are exponential, the hyper-parameters $(\alpha, \beta)$ and $(\alpha', \beta')$ are not necessarily identical. In a later section (3.2) we will utilize the concept of relaxation time shift factors (Section 2.2.2) to relate the complex modulus parameters obtained from the elastic experiments to corresponding changes in the full viscoelastic properties of the material.

### 3.1.3 Multi-body compound effect

The compound effect describes the phenomenon that when two filler aggregates are close to each other, their interphases could interact providing additional enhancement of the property of the interphase area between the aggregates. The work by Zhang et al. [35] quantitatively demonstrates this effect and in this work we use data from the 520 nm and 156nm sample to mathematically describe the multi-body compound effect.

As per [35], the filler substrates in the 520nm sample is considered to be far enough to avoid interphase interaction, while the 156nm sample is observed to have increased interphase property incurred by interphase interaction. Therefore, Eqn. 7 is first used to learn the single body interphase gradient from the 520nm sample, defined as $\overline{E_d} = F(d)$ as in the previous section. For the 156 nm sample, for a location which has a distance of $d_1$ from the left substrate, the distance from the right substrate is $(156 - d_1)$ nm. At this location, the theoretical normalized elastic moduli affected by either side are $\overline{E_{d_1}} = F(d_1)$ and $\overline{E_{d_2}} = F(d_2) = F(156 - d_1)$ respectively. At the same time, using the data from the 156nm sample and Eqn. 7, an exponential decay function (shown in **Fig. 5**) $\overline{E_d^{compound}} = \overline{E_d^{156nm}} = H(d) = G(F(d))$ is also obtained. It is noted that $H(d)$ reflects the observed experimental effects of single-body interphase gradient $F(\cdot)$ and two-body compound effect $G(\cdot)$, and it is directly dependent on the distance d from the substrate surface in the AFM sample. In contrast, $G(\cdot)$ is the mathematical function of compound effect that is not observable from experiments and is not distance dependent. Because of the symmetry of the substrate-polymer-substrate model composite sample, it is straightforward to take $H(d)$ as the compound property when $0 < d \leq 78nm$, and take $H(156 - d)$ when $78 < d < 156nm$. By varying $d_1$, a dataset $\left(\overline{E_{d_1}}, \overline{E_{d_2}}, \overline{E_{d_1}^{compound}}\right) = \left(\overline{E_{d_1}}, \overline{E_{d_2}}, \overline{E_{d_1}^{156nm}}\right)$ is collected and the compound effect $G(\cdot)$ could be learned by $\overline{E_{d_1}^{compound}} = G(\overline{E_{d_1}}, \overline{E_{d_2}})$.

While the compound effect is learned from the AFM experimental data in which only two filler substrates are included, it is desired to develop a generalized form for describing the interphase impacted by multiple (n>2) filler aggregates. In this generalization, the functional form of the compound effect $G(\cdot)$ (Eqn. 3) should satisfy some constraints: 1) the functional form has to be symmetric with respect to $\overline{E_{d_1}}, \overline{E_{d_2}}, \ldots, \overline{E_{d_n}}$. Otherwise, interchanging $\overline{E_{d_i}}$ and $\overline{E_{d_j}}$ ($i \neq j$) would result in different compound properties. 2). The functional form should be monotonically decreasing with respect to distances $d_i$ from aggregates. 3). The inclusion of the single-body interphase gradient $\overline{E_{d_i}}$ created by aggregate #i should lead to the enhancement of compound property. Mathematically, this latter constraint is expressed as,

$$G(|\overline{E_{d_1}}|, |\overline{E_{d_2}}|) < G(|\overline{E_{d_1}}|, |\overline{E_{d_2}}|, |\overline{E_{d_3}}|) < G(|\overline{E_{d_1}}|, |\overline{E_{d_2}}|, \ldots, |\overline{E_{d_n}}|) \quad (8)$$

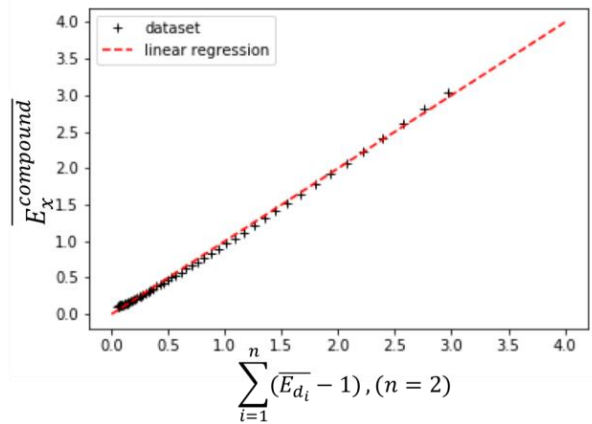

**Figure 6 Linear regression for functional form of compound effect**

Several mathematical forms that satisfy these three constraints (e.g. the sum or the sum of the squares of the normalized elastic modulus) are extracted and the relationship between these features and the compound property is analyzed. A linear relationship (**Fig. 6**) between the compound normalized elastic modulus and the term $\sum_{i=1}^{n}(\overline{E_{d_i}} - 1)$ is identified as,

$$\overline{E_x^{compound}} - 1 = \eta'[\sum_{i=1}^{n}(\overline{E_{d_i}} - 1)] + \xi' \qquad (9)$$

where the location $x$ determines the surface distances $d_i$, while $\eta' = 1.08$, $\xi' = 0.06$ and $n = 2$ for the fitted regression. Conceptually, $(\overline{E_{d_i}} - 1)$ represents the virtual enhancement of elastic modulus produced by filler aggregate #i, and ($\overline{E_x^{compound}} - 1$) indicates the combined enhancement of the elastic modulus above the matrix property by all the surrounding filler aggregates. $\eta'$ is a parameter describing the interactive behavior between interphases created by different aggregates and $\xi'$ is a compensating factor to correct the under/over-estimation. By analogy, the compound effect for viscoelastic property is assumed to be,

$$\overline{|E_{compound,f}^*|} - 1 = \eta[\sum_{i=1}^{n}(\overline{|E_{d_i,f}^*|} - 1)] + \xi \qquad (10)$$

With this formulation, the parameter set $\mathbf{\Omega}$ in Eqn. 3 is specified by $\mathbf{\Omega} = (\eta, \xi)$.

### 3.2 Determination of the shifting factors at each location in the interphase

To implement the FE simulations of the composite properties using a gradient interphase concept, we must determine the shifting factor at each location of the polymer material. This shifting factor at any location in the interphase is determined by: 1) identifying the number of aggregates n that may affect this location by specifying a sufficiently large cut-off distance. 2). For each filler aggregate #i (i=1…N), computing the theoretical single-body interphase gradient $\overline{|E^*_{d_i,f}|}$ using Eqn. 8, 3). Applying the compound effect (Eqn. 11) to combine $\overline{|E^*_{d_i,f}|}$ to obtain $|E^*_{compound,f}|$, and 4) as illustrated in **Fig. 7**, $\overline{|E^*_{compound,f}|}$ is essentially the vertical enhancement of the normalized magnitude of complex modulus at frequency $f$. The corresponding shifting factor, $S_{intph}$, can then be identified according to $\overline{|E^*_{compound,f}|}$ as illustrated in **Fig. 7**.

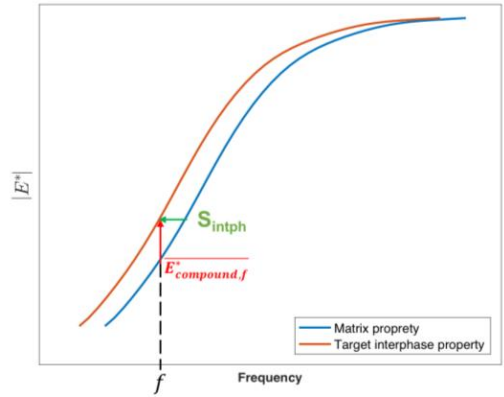

**Figure 7 An illustration of the conversion between the normalized magnitude of complex modulus and shifting factor for interphase. $\overline{|E^*_{compound,f}|}$ is the normalized magnitude of complex modulus at frequency estimated by multi-body compound effect.**

### 3.3 Numerical studies of the interphase effects

In the prior sections, the proposed interphase representation is first presented, followed by a data-driven exploration of the mathematical forms of the representations. In this section, we first explore the parametric influences of the hyper-parameters in the proposed interphase representations. After that, we study the equivalent spatial distributions of the interphase representations for a prior numerical model with uniform interphase assumption, by applying a Bayesian Inference method. Last, the proposed interphase representation is applied to describe the spatial distribution of interphase property of an experimental sample of polymer composite.

#### 3.3.1 The effects of single-body interphase gradient

The effects of single-body interphase gradient are first investigated. Using the circle packing algorithm in [44], a dilute microstructure (**Fig. 8(a)**) is generated to avoid the interaction between interphases created by different filler aggregates – in this case only the single interphase effect will be relevant around each particle. The values of the two parameters $\alpha$ and $\beta$ from Eqn. 8 in the single-body interphase gradient are altered individually and the comparisons between the simulated viscoelastic properties by altering $\alpha$ and $\beta$ are shown in **Fig. 9(a) & (b).** It is observed that $\beta$, the term in the exponent in Eqn. 8, has greater effects on the bulk property than $\alpha$. This effect is illustrated visually in **Fig. 9(c) & (d)**, where the distributions of shifting factors are shown. The contour plots of shift factor magnitude demonstrate that increasing the value of $\alpha$ only affects the property of the interphase regions closely surrounding the filler aggregate, and thus causing only small differences from the bulk composite property. In contrast, varying the $\beta$ value can substantially extend the interphase region (**Fig. 9(d)**) and more significantly affect the bulk composite property.

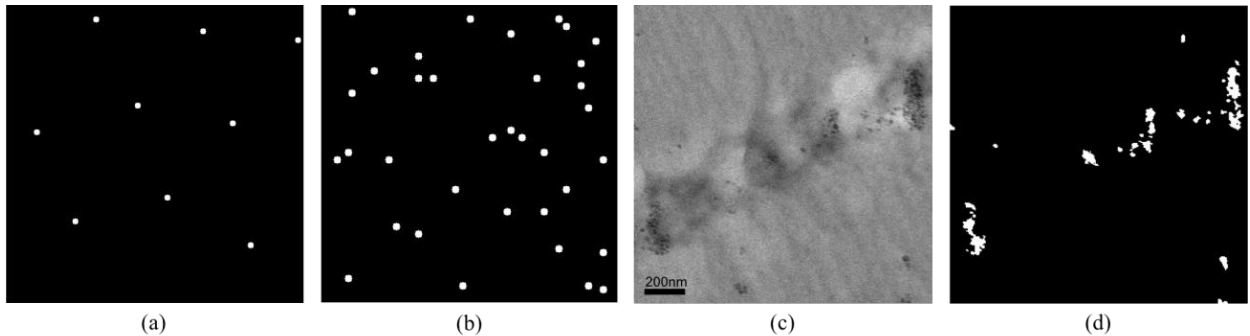

Figure 8 Microstructures used for studying the effects of interphase representation. (a) a dilute microstructure (VF=0.29%) for studying single-body interphase gradient, (b) a moderately loaded microstructure (VF=1.77%) for studying multi-body compound effect, (c) Transmission Electron Microscopy (TEM) image of Polystyrene-silica composite, and (d) the binarized image (VF=1.83%) of (c) using Niblack algorithm[45, 46].

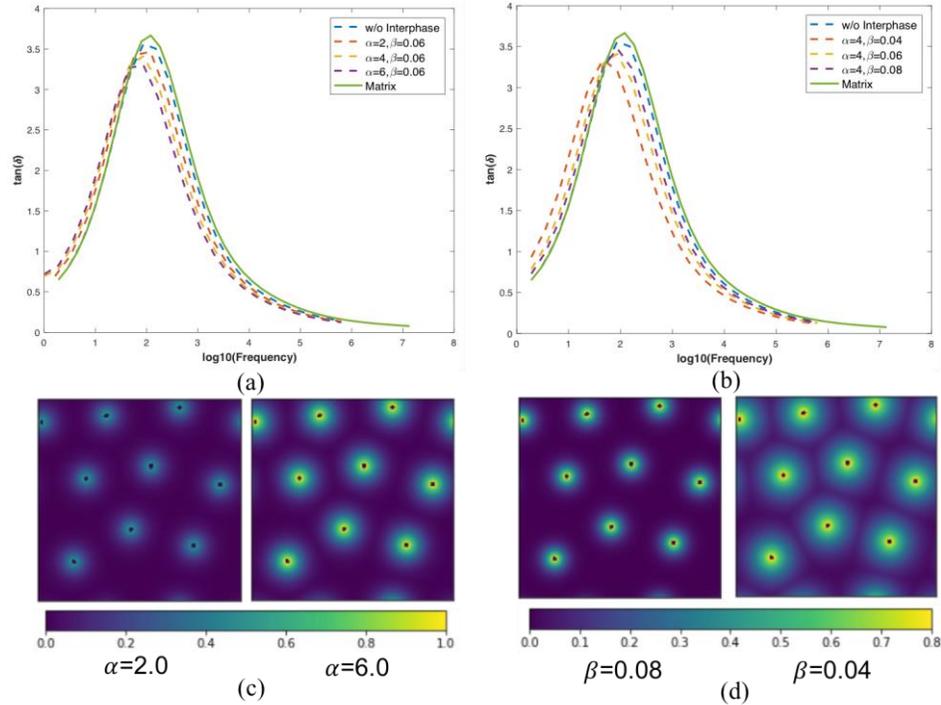

**Figure 9** Effects of single-body interphase gradient. (a) the comparison between the simulated viscoelastic properties by different $\alpha$, (b) the comparison between the simulated viscoelastic properties by different $\beta$, (c) the distribution of shifting factor magnitude when altering $\alpha$, and (d) the change of shifting factor distribution when altering $\beta$. The color map represents the value of the shift factor, S (see Figure 3); e.g. in (c) a shift of 1 decade from the matrix properties is reflected by a color value of 1.

### 3.3.2 The effects of multi-body interphase compound effect

With the effects of single-body interphase gradient as a baseline, the influence of multi-body interphase compound effect is also investigated. In this numerical study, a moderately loaded microstructure (**Fig. 8(b)**) is generated and different settings of interphase compound effect are tested. Specifically, the generalization of compound effect from two filler AFM experimental samples to multiple filler geometries (discussed in Section 3.1) is applied to allow the interaction between multiple filler aggregates. The cut-off radius for considering a specific location impacted by a filler is set to 100 pixels for this example (half length of the microstructure), while the exponential form of the single-body interphase gradient (Eqn. 7 &8) indicates that only the closest particle will have significant impact to the property at a given material point. The use of a large potential influence window (defined by the cut-off distance) in the algorithm is more general and allows the interaction functions to effectively handle the level of influence of closer and more distant particles. **Fig. 10 (a) & (b)** show the comparisons between the simulated composite properties with different $\eta, \xi$ values. **From Fig. 10(a) & (b),** it is found that increasing $\eta$ could lower the magnitude of the $\tan(\delta)$ peak and shift it to lower frequency. In contrast, altering $\xi$ shifts the $\tan(\delta)$ curve does not significantly change its magnitude. This effect is visually illustrated by viewing the shift factors in Fig. **10(c) & (d)**: (c) shows that increasing $\eta$ enhances the interactions between interphases, which affects the $\tan(\delta)$ curve in both horizontal and vertical directions. In comparison, changing $\xi$ results in an increment of shifting factor magnitude, which leads to a shifting of the bulk property.

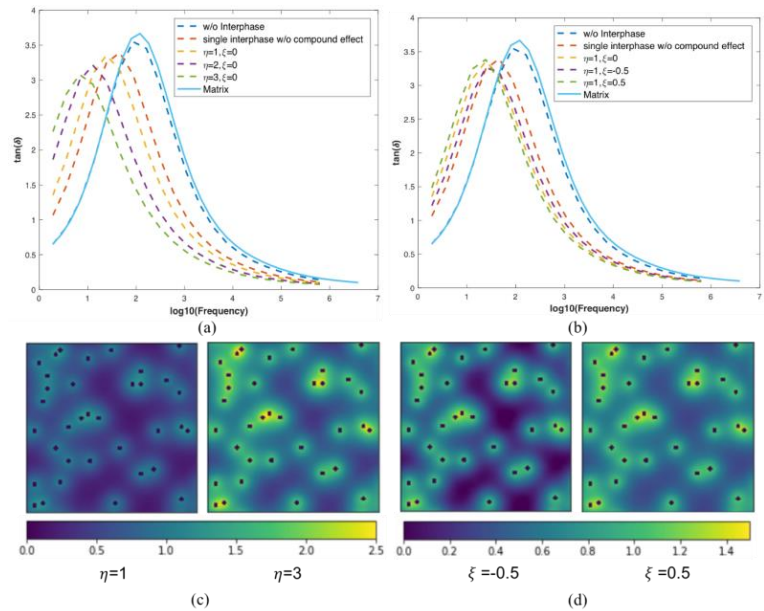

**Figure 10. Effects of multi-body interphase compound effect. (a) the comparison between the simulated viscoelastic properties by different $\eta$, (b) the comparison between the simulated viscoelastic properties by different $\xi$, (c) the distribution of shifting factor magnitudes when altering $\eta$ and (d) the distribution of shifting factor magnitudes when altering $\xi$. The color map represents the value of the shift factor, S (see Figure 3); e.g. in (c) a shift of 1 decade from the matrix properties is reflected by the color value indicated for 1.**

### 3.3.3 Numerical comparison to uniform interphase modeling

In our prior work [12], without the access to the experimental measurement of local mechanical property, it was assumed that the interphase property was related to the matrix property in frequency domain by a simple two decade shift in relaxation times. In this section, a comparative study is conducted to demonstrate that the proposed gradient interphase representation can not only predict the bulk property of the composite as the prior one but also supply additional interphase information. First, a microstructure of Polystyrene(PS)-Silica composite is chosen (**Fig. 8(d)**) and the master curve of PS referenced at 120C degree is utilized as the matrix property. We assume that the functional forms of the interphase representations that we learned from earlier analysis is applicable to this example. The target bulk viscoelastic property is simulated by the prior uniform interphase model, in which interphase property is assumed to be shifted from matrix property by 2 decades. In this simulation, the thickness of interphase is set as 30 pixels (146nm, while the particle diameters are 40nm on average). Second, we iteratively run the FE model with gradient interphase under the Bayesian Inference framework to achieve equivalence of bulk properties predicted using a gradient interphase with those from the target uniform interphase case. In these simulations, 100Hz is used as the reference frequency $f$ in Eqn. 8 and Eqn. 11 for estimating the shifting factor for each location in the interphase. In addition, the cut-off distance for considering the impact from a filler is again set as 100 pixels (~480nm) and the ranges for the hyper-parameters are specified as $\alpha \in [4, 6.5]$, $\beta \in [0.06, 0.1]$, $\eta \in [0.8, 1.2]$ and $\xi \in [-0.5, 0.5]$ and the objective function is set as,

$$\varepsilon = \sum_{i=1}^{m}[\tan(\delta)_{uniform} - \tan(\delta)_{gradient}]^2_{f=f_i} \quad (12)$$

where $f_i$s are the frequencies that the viscoelastic property is evaluated at, and the number of frequencies to evaluate, $m$, is set as 30 in this numerical study.

Then the values of ($\Phi, \Omega$) that minimize $\varepsilon$ are identified using the Bayesian Inference approach for 50 iterations, as illustrated in Section 2.4.

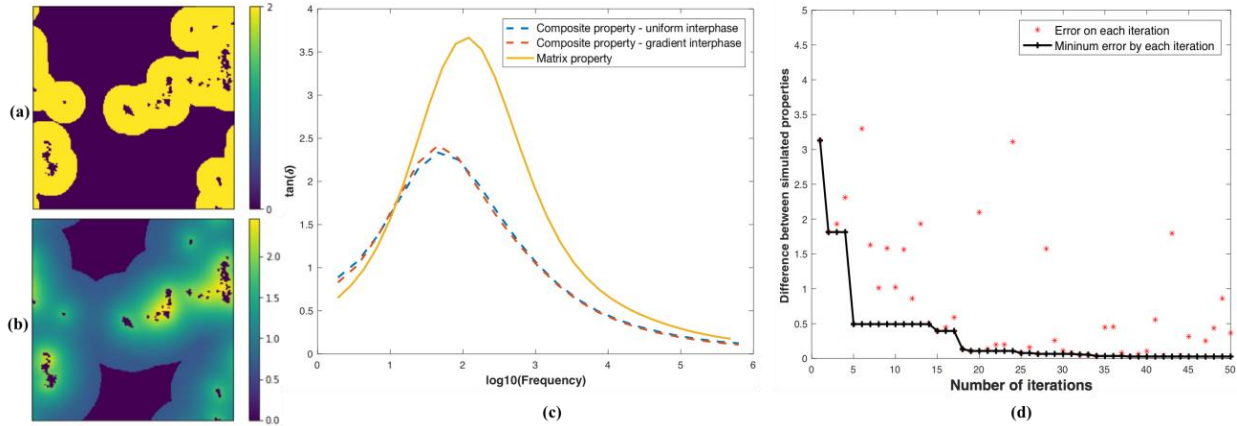

**Figure 11 The comparison between the FE models using uniform interphase and gradient interphase via Bayesian Inference. (a) the distribution of shifting factor magnitude for the uniform interphase FE model. The interphase property is uniformly distributed and shifted from matrix property by 2 decades, (b) the distribution of shifting factor magnitudes for the proposed gradient interphase FEA model, (c) the comparison between the matrix property and the simulated properties of the two models, and (d) the history of Bayesian Inference.**

After the Bayesian Inference, it is found that the values of ($\alpha, \beta, \eta, \xi$) to achieve the model equivalence (error = 0.03) are (4.22, 0.09, 0.80, -0.50) respectively. Key observations from these simulations are: 1) compared to the previous uniform interphase model (**Fig. 11(a)**), the proposed gradient interphase representation **(Fig. 11(b))** can effectively model the interphase property with physically realistic gradients, while achieving the same bulk composite property (**Fig. 11(c)**). 2) The Bayesian Inference approach is capable of exploring the optimal solution efficiently and reducing the number of FE simulations (**Fig. 11(d)).** 3). From **Fig. 11(b)**, it is found that the interphase areas that are affected by many filler aggregates are significantly strengthened. 4) Using Eqn. 8 and the criteria of interphase thickness determination (105% of the matrix property) describe in Section 2.2, the interphase thickness in the proposed model is obtained as 54.84 pixels. The total interphase thickness in the gradient interphase case is larger than that of the uniform interphase model because single-body interphase gradient drops exponentially as the distance increases, naturally extending its domain.

### 3.3.4 Numerical Verification with Experimental Data

In Section 3.3.3, the prior uniform interphase modeling approach is utilized to produce a virtual objective of the material viscoelastic property, and the equivalent gradient interphase representation is learned via Bayesian Inference. In this section we demonstrate the descriptive capability of the proposed interphase representation by directly utilizing experimental data for a composite in [47] and investigate how the proposed gradient interphase representation could describe the distribution of local properties within the microstructure.

In this demonstration, the microstructure of the composite (Fig. 12(a)) is extracted and binarized (Fig. 12(b)). Meanwhile, the temperature sweeps of the viscoelastic properties of the matrix and composite in that work are also converted to frequency dependent spectrums (the green and the blue curves in Fig. 12(d)) using Willams-Landel-Ferry equation[48]. These data, together with the identified mathematical forms of the proposed interphase gradient representations in Section 3.1, is then fed into the Bayesian Inference framework discussed in Section 2.3 to match the simulated bulk property to the experimental data. After the Bayesian Interference computations, the values of the hyper-parameters in the functional forms of the interphase representations are identified and the spatial distribution of local properties within the microstructure can be visualized. It is also noted that, since the microstructure in this numerical validation has different length scale from the previous one used in Section 3.3.3, the ranges of the hyper-parameters in the Bayesian optimization are adjusted accordingly as $\alpha \in [1.5, 6.5]$, $\beta \in [0.12, 0.25]$, $\eta \in [0.5, 1,1]$ and $\xi \in [-1.5, 1.0]$, and Eqn. 12 is kept as the objective function.

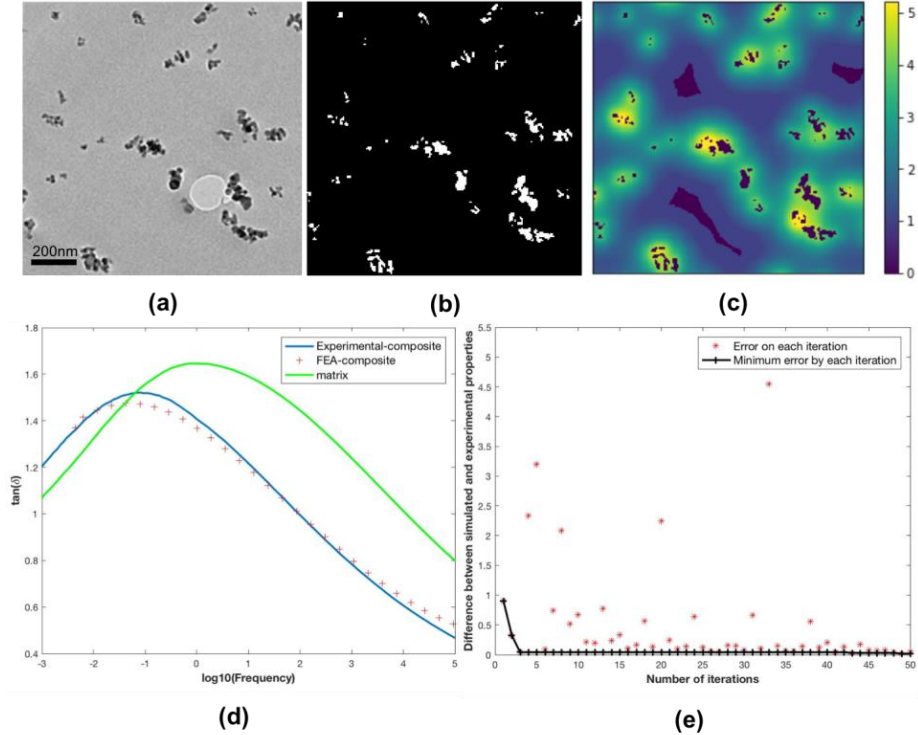

**Figure 12. The results of numerical validation of the proposed gradient interphase representation on experimental data from [47]. (a) Microstructure image gathered from [47] (b) binary image of the microstructure. (c) the distribution of shifting factor magnitudes for the proposed gradient interphase FEA model. (d) the comparison between the matrix property, experimental measurements of the composite property and the simulated property. (e) the history of Bayesian Inference.**

As shown in **Figure 12(d),** after employing the Bayesian Inference method, the simulated property of the composite agrees well with the experimental measurements (error = 0.04). The corresponding values of the hyper-parameters are $(\alpha, \beta, \eta, \xi) = (2.37, 0.15, 0.5, 0.0185)$. **Figure 12** demonstrates that 1). the proposed FE model can effectively model the spatial distribution of the local properties in the microstructure (**Figure 12(c)**), while matching with the experimental viscoelastic property data. 2). The Bayesian Inference framework can efficiently explore the space of the hyper-parameters and identify these values of hyper-parameters. 3) Using Eqn. 7 and the criteria of interphase thickness determination (105% of the matrix property) describe in Section 2.2, the interphase thickness in the proposed model is obtained as 26.10 pixels, which corresponds to 166nm. While this interphase thickness identified in this work is of the same magnitude as found in our prior work (~100 nm in [34]), it is slightly greater than that in [34]. This finding could be due to 1) the difference between elastic and viscoelastic properties: the interphase thickness in [34] was inferred using elastic property while in this work the magnitude of complex modulus is utilized. Different properties could perform slightly differently in reflecting the interphase thickness. 2) Numerical error introduced by coarsened meshing: in our FE model, the original microstructure which has a resolution of 700x700 pixels is coarsened to a voxelated meshing (200x200) for computational efficiency, and this coarsening process may introduce numerical errors.

It should be also noted that while the functional form of the compound effect is learned based on the experimental data with two overlapped interphases (Section 3.1.2), our method can handle the interaction between interphases produced by multiple particle aggregates. **Figure 13** illustrates the distribution of the number of interacting interphases in the microstructure for the two validations conducted in this work. It is observed that in the region where many particle aggregates are densely distributed, it is possible that a particular location is affected by more than 10 nearby aggregates. This again verifies the capability of our method in handling multi-body (greater than 2) compound effect and complex microstructure dispersions.

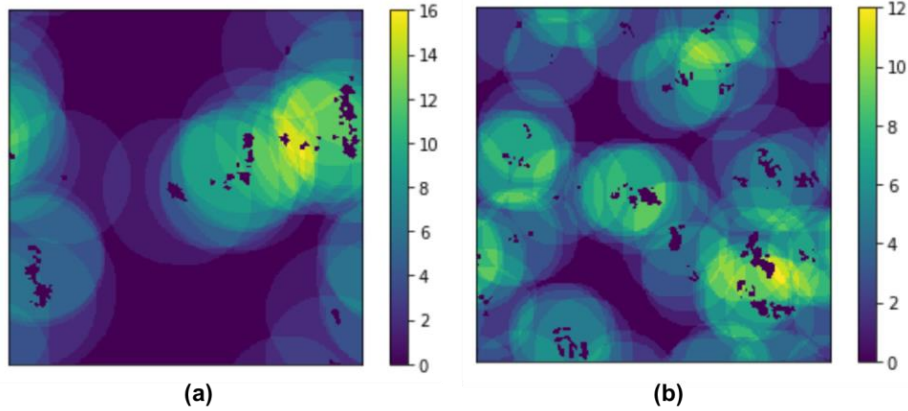

**Figure 13. The distributions of the number of overlapped interphases in the two test cases of this work. (a) & (b) correspond to the microstructures in Figure 11 & 12 respectively**.

## 4  Conclusion and Future Work

Based upon AFM experimental observations of local polymer interphase properties, we proposed here a new gradient interphase representation. The proposed interphase representation contains two components, single-body interphase gradient, which represents the effects of single filler aggregates, and multi-body compound effect, which describes the interaction between interphases created by multiple filler aggregates. The proposed interphase representation is implemented in Finite Element simulation and numerical studies are conducted to investigate the effects of each interphase representation components. In addition, the proposed interphase representation is compared numerically with previously developed uniform interphase model via FE modeling and Bayesian Inference. It is demonstrated that the proposed interphase can achieve the equivalent property prediction capability as the prior method, and also supply more reasonable interphase gradient information. In addition to the numerical validation on virtual material property, it is also shown that the proposed interphase representation could facilitate the exploration of spatial property distribution within the interphase by utilizing Bayesian Inference on experimental measurements.

While this work presents the complete process of utilizing AFM experimental measurements of local interphase properties in inferring spatial distribution of interphase via Finite Element modeling, several potential directions could be pursued to further extend this work. First, in this work, the mathematical form of compound effect is learned from the AFM experimental data in which only two filler substrates are included, and the generalization of the compound effect for multiple (n>2) interphase interaction is not rigorously validated. AFM model composites with a great variance of complex geometries, if available, would reveal the impacts of microstructural descriptors for multiple interphase cases, and they would potentially lead to a more precise functional form of the compound effect. Then this sophisticated interphase representations could be utilized to predict the properties of given composites in addition to the inference of interphase properties presented in this work. Second, while the discussion in this work is limited to modeling the viscoelastic property of polymer nanocomposites, similar interphase representation could also be developed in the similar manner for other material properties of interest (e.g. dielectric property if local dielectric property within interphase is available). Lastly, the proposed interphase representation could be implemented in 3D FE modeling.

## 5 Acknowledgment

This work is supported by Center of Hierarchical Materials Design (NIST CHiMad 70NANB14H012), NSF-DMREF (NSF Award 1818574, 1729743, 1729452) and NSF-DIBBS (NSF Award 1640840, A12761, 1640840).

**Reference**

1. Brinson, H.F. and L.C. Brinson, *Characteristics, applications and properties of polymers*, in *Polymer Engineering Science and Viscoelasticity*. 2015, Springer. p. 57-100.
2. Schadler, L., L. Brinson, and W. Sawyer, *Polymer nanocomposites: a small part of the story.* Jom, 2007. **59**(3): p. 53-60.
3. Koo, J.H., *Polymer nanocomposites*. 2006: McGraw-Hill Professional Pub.
4. Thomason, J., *Investigation of composite interphase using dynamic mechanical analysis: artifacts and reality.* Polymer composites, 1990. **11**(2): p. 105-113.
5. Ray, S.S. and M. Okamoto, *Polymer/layered silicate nanocomposites: a review from preparation to processing.* Progress in polymer science, 2003. **28**(11): p. 1539-1641.
6. DiBenedetto, A., *Tailoring of interfaces in glass fiber reinforced polymer composites: a review.* Materials Science and Engineering: A, 2001. **302**(1): p. 74-82.
7. Winey, K.I. and R.A. Vaia, *Polymer nanocomposites.* MRS bulletin, 2007. **32**(4): p. 314-322.
8. Gass, J., et al., *Superparamagnetic polymer nanocomposites with uniform Fe3O4 nanoparticle dispersions.* Advanced Functional Materials, 2006. **16**(1): p. 71-75.
9. Zou, H., S. Wu, and J. Shen, *Polymer/silica nanocomposites: preparation, characterization, properties, and applications.* Chem. Rev, 2008. **108**(9): p. 3893-3957.
10. Fu, X. and S. Qutubuddin, *Polymer–clay nanocomposites: exfoliation of organophilic montmorillonite nanolayers in polystyrene.* Polymer, 2001. **42**(2): p. 807-813.
11. Chen, G.H., et al., *Preparation of polymer/graphite conducting nanocomposite by intercalation polymerization.* Journal of Applied Polymer Science, 2001. **82**(10): p. 2506-2513.
12. Qiao, R. and L.C. Brinson, *Simulation of interphase percolation and gradients in polymer nanocomposites.* Composites Science and Technology, 2009. **69**(3-4): p. 491-499.
13. Moniruzzaman, M. and K.I. Winey, *Polymer nanocomposites containing carbon nanotubes.* Macromolecules, 2006. **39**(16): p. 5194-5205.
14. Seiler, J. and J. Kindersberger. *Polymer-filler interactions and polymer chain dynamics in the interphase in silicon nanocomposites*. in *Dielectrics (ICD), 2016 IEEE International Conference on*. 2016. IEEE.
15. Subramanian, R. and A. Crasto, *Electrodeposition of a polymer interphase in carbon-fiber composites.* Polymer composites, 1986. **7**(4): p. 201-218.
16. Gatenholm, P., H. Bertilsson, and A. Mathiasson, *The effect of chemical composition of interphase on dispersion of cellulose fibers in polymers. I. PVC-coated cellulose in polystyrene.* Journal of applied polymer science, 1993. **49**(2): p. 197-208.
17. Ghanbari, A., et al., *Interphase structure in silica–polystyrene nanocomposites: a coarse-grained molecular dynamics study.* Macromolecules, 2011. **45**(1): p. 572-584.
18. Ciprari, D., K. Jacob, and R. Tannenbaum, *Characterization of polymer nanocomposite interphase and its impact on mechanical properties.* Macromolecules, 2006. **39**(19): p. 6565-6573.
19. Hashin, Z., *Analysis of composite materials—a survey.* Journal of Applied Mechanics, 1983. **50**(3): p. 481-505.
20. Christensen, R.M., *A critical evaluation for a class of micro-mechanics models*, in *Inelastic Deformation of Composite Materials*. 1991, Springer. p. 275-282.

21. Mori, T. and K. Tanaka, *Average stress in matrix and average elastic energy of materials with misfitting inclusions.* Acta metallurgica, 1973. **21**(5): p. 571-574.
22. Fisher, F. and L. Brinson, *Viscoelastic interphases in polymer–matrix composites: theoretical models and finite-element analysis.* Composites Science and technology, 2001. **61**(5): p. 731-748.
23. Benveniste, Y., G. Dvorak, and T. Chen, *Stress fields in composites with coated inclusions.* Mechanics of Materials, 1989. **7**(4): p. 305-317.
24. Deng, F. and K.J. Van Vliet, *Prediction of elastic properties for polymer–particle nanocomposites exhibiting an interphase.* Nanotechnology, 2011. **22**(16): p. 165703.
25. Ji, X.L., et al., *Tensile modulus of polymer nanocomposites.* Polymer Engineering & Science, 2002. **42**(5): p. 983-993.
26. Boutaleb, S., et al., *Micromechanics-based modelling of stiffness and yield stress for silica/polymer nanocomposites.* International Journal of Solids and Structures, 2009. **46**(7-8): p. 1716-1726.
27. Zhu, L. and K. Narh, *Numerical simulation of the tensile modulus of nanoclay-filled polymer composites.* Journal of Polymer Science Part B: Polymer Physics, 2004. **42**(12): p. 2391-2406.
28. Read, D., et al., *Theoretical and finite-element investigation of the mechanical response of spinodal structures.* The European Physical Journal E, 2002. **8**(1): p. 15-31.
29. Qiao, R., et al., *Effect of particle agglomeration and interphase on the glass transition temperature of polymer nanocomposites.* Journal of Polymer Science Part B: Polymer Physics, 2011. **49**(10): p. 740-748.
30. Hu, A., et al., *Computational analysis of particle reinforced viscoelastic polymer nanocomposites–statistical study of representative volume element.* Journal of the Mechanics and Physics of Solids, 2018. **114**: p. 55-74.
31. Xu, H., et al., *Stochastic reassembly strategy for managing information complexity in heterogeneous materials analysis and design.* Journal of Mechanical Design, 2013. **135**(10): p. 101010.
32. Wang, Y., et al., *Identifying interphase properties in polymer nanocomposites using adaptive optimization.* Composites Science and Technology, 2018.
33. Downing, T., et al., *Determining the interphase thickness and properties in polymer matrix composites using phase imaging atomic force microscopy and nanoindentation.* Journal of adhesion science and technology, 2000. **14**(14): p. 1801-1812.
34. Cheng, X., et al., *Characterization of local elastic modulus in confined polymer films via AFM indentation.* Macromolecular rapid communications, 2015. **36**(4): p. 391-397.
35. Zhang, M., et al., *Stiffness Gradients in Glassy Polymer Model Nanocomposites: Comparisons of Quantitative Characterization by Fluorescence Spectroscopy and Atomic Force Microscopy.* Macromolecules, 2017. **50**(14): p. 5447-5458.
36. Xu, H., et al., *Descriptor-based methodology for statistical characterization and 3D reconstruction of microstructural materials.* Computational Materials Science, 2014. **85**: p. 206-216.
37. Wang, D., et al., *Visualization of nanomechanical mapping on polymer nanocomposites by AFM force measurement.* Polymer, 2010. **51**(12): p. 2455-2459.
38. Kolluru, P.V., et al., *An AFM-based Dynamic Scanning Indentation (DSI) Method for Fast, High Resolution Spatial Mapping of Local Viscoelastic Properties in Soft Materials.* Macromolecules (in press), 2018.

39. Nakajima, K., et al., *Nanorheology of polymer blends investigated by atomic force microscopy.* Japanese journal of applied physics, 1997. **36**(6S): p. 3850.
40. Deng, H., et al., *Utilizing real and statistically reconstructed microstructures for the viscoelastic modeling of polymer nanocomposites.* Composites Science and Technology, 2012. **72**(14): p. 1725-1732.
41. Zhao, H., et al., *Dielectric spectroscopy analysis using viscoelasticity-inspired relaxation theory with finite element modeling.* IEEE Transactions on Dielectrics and Electrical Insulation, 2017. **24**(6): p. 3776-3785.
42. Yang, Z., et al., *Microstructural Materials Design via Deep Adversarial Learning Methodology.* arXiv preprint arXiv:1805.02791, 2018.
43. Li, X.Y., Zijiang; Brinson, L Catherine; Choudhary, Alok N; Agrawal, Ankit; Chen, Wei, *A deep adversarial learning approach for designing microstructural material systems*, in *The ASME 2018 International Design Engineering Technical Conferences & Computers and Information in Engineering Conference (IDETC/CIE 2018)*. 2018: Quebec City, Quebec, Canada.
44. Yu, S., et al., *Characterization and design of functional quasi-random nanostructured materials using spectral density function.* Journal of Mechanical Design, 2017. **139**(7): p. 071401.
45. Hassinger, I., et al., *Toward the development of a quantitative tool for predicting dispersion of nanocomposites under non-equilibrium processing conditions.* Journal of materials science, 2016. **51**(9): p. 4238-4249.
46. Silver, B., *An introduction to digital image processing.* Cognex Corporation, 2000.
47. Saladino, M., et al., *The effect of silica nanoparticles on the morphology, mechanical properties and thermal degradation kinetics of PMMA.* Polymer degradation and stability, 2012. **97**(3): p. 452-459.
48. Williams, M.L., R.F. Landel, and J.D. Ferry, *The temperature dependence of relaxation mechanisms in amorphous polymers and other glass-forming liquids.* Journal of the American Chemical society, 1955. **77**(14): p. 3701-3707.